# Overcoming the Challenges Associated with Image-based Mapping of Small Bodies in Preparation for the OSIRIS-REx Mission to (101955) Bennu

D. N. DellaGiustina[1], C. A. Bennett[1], K. Becker[1], D. R Golish[1], L. Le Corre[2], D. A. Cook[3†], K. L. Edmundson[3], M. Chojnacki[1], S. S. Sutton[1], M. P. Milazzo[3], B. Carcich[4], M. C. Nolan[1], N. Habib[1], K. N. Burke[1], T. Becker[1], P. H. Smith[1], K. J. Walsh[5], K. Getzandanner[6], D. R. Wibben[4], J. M. Leonard[4], M. M. Westermann[1], A. T. Polit[1], J. N. Kidd Jr.[1], C. W. Hergenrother[1], W. V. Boynton[1], J. Backer[3], S. Sides[3], J. Mapel[3], K. Berry[3], H. Roper[1], C. Drouet d'Aubigny[1], B. Rizk[1], M. K. Crombie[7], E. K. Kinney-Spano[8], J. de León[9, 10], J. L. Rizos[9, 10], J. Licandro[9, 10], H. C. Campins[11], B. E. Clark[12], H. L. Enos[1], and D. S. Lauretta[1]

[1]Lunar and Planetary Laboratory, University of Arizona, Tucson, AZ, USA

[2]Planetary Science Institute, Tucson, AZ, USA

[3]U.S. Geological Survey Astrogeology Science Center, Flagstaff, AZ, USA

[4]KinetX Space Navigation & Flight Dynamics Practice, Simi Valley, CA, USA

[5]Southwest Research Institute, Boulder, CO, USA

[6]Goddard Spaceflight Center, Greenbelt, MD, USA

[7]Indigo Information Services LLC, Tucson, AZ, USA

[8] MDA Systems, Ltd, Richmond, BC, Canada

[9]Instituto de Astrofísica de Canarias, Santa Cruz de Tenerife, Spain

[10]Departamento de Astrofísica, Universidad de La Laguna, Santa Cruz de Tenerife, Spain

[11]Department of Physics, University of Central Florida, Orlando, FL, USA

[12]Department of Physics and Astronomy, Ithaca College, Ithaca, NY, USA

[†]Retired from this institution

Corresponding author: Daniella N. DellaGiustina (danidg@lpl.arizona.edu)

**Key Points:**

- The OSIRIS-REx Asteroid Sample Return Mission performs image-based mapping of (101955) Bennu to aid in the selection of a sample-site

- We develop observational strategies to perform mapping to address the challenges associated with surveying a small body

- We identify pitfalls and best practices for mapping images of small bodies with large concavities, elongated axes, or overhanging terrain

## Abstract

The OSIRIS-REx Asteroid Sample Return Mission is the third mission in NASA's New Frontiers Program and is the first U.S. mission to return samples from an asteroid to Earth. The most important decision ahead of the OSIRIS-REx team is the selection of a prime sample-site on the surface of asteroid (101955) Bennu. Mission success hinges on identifying a site that is safe and has regolith that can readily be ingested by the spacecraft's sampling mechanism. To inform this mission-critical decision, the surface of Bennu is mapped using the OSIRIS-REx Camera Suite and the images are used to develop several foundational data products. Acquiring the necessary inputs to these data products requires observational strategies that are defined specifically to overcome the challenges associated with mapping a small irregular body. We present these strategies in the context of assessing candidate sample-sites at Bennu according to a framework of decisions regarding the relative safety, sampleability, and scientific value across the asteroid's surface. To create data products that aid these assessments, we describe the best practices developed by the OSIRIS-REx team for image-based mapping of irregular small bodies. We emphasize the importance of using 3D shape models and the ability to work in body-fixed rectangular coordinates when dealing with planetary surfaces that cannot be uniquely addressed by body-fixed latitude and longitude.

## Plain Language Summary

The OSIRIS-REx Asteroid Sample Return Mission must map asteroid (101955) Bennu using the OSIRIS-REx Camera Suite. Here we present the techniques that are established to accomplish this goal. Mapping helps us find the best place on the surface of Bennu from which to gather a sample. Because asteroids are small bodies with weak gravitational fields, maneuvering a spacecraft around them can be challenging. Considering these complexities, we have found ways to gather images of Bennu needed for creating maps. Additionally, due to the irregular shape of many asteroids, producing 2D maps in terms of latitude and longitude may be insufficient for describing their surface geography. To that end, we have developed software that is capable of creating and displaying image maps in 3D.

## 1 Introduction

The aim of NASA's Origins, Spectral Interpretation, Resource Identification, Security-Regolith Explorer (OSIRIS-REx) mission is to study the asteroid (101955) Bennu (Lauretta et al., 2015) and return a pristine sample of its surface material to Earth for detailed characterization. Prior to the sampling event, the OSIRIS-REx spacecraft surveys Bennu for approximately two years using a suite of instruments to select and document the most eminent candidate sample-sites. The auxiliary role of remote sensing in support of the primary objective of sample return makes OSIRIS-REx somewhat unique among planetary missions. By combining coordinated observations from the OSIRIS-REx Camera Suite (OCAMS; Rizk et al., 2018), the OSIRIS-REx Laser Altimeter (OLA; Daly et al., 2017), the OSIRIS-REx Visible and InfraRed Spectrometer (OVIRS; Reuter et al., 2018), the OSIRIS-REx Thermal Emission Spectrometer (OTES; Christensen et al., 2018), the REgolith X-Ray Imaging Spectrometer (REXIS; Masterson et al., 2018), and the OSIRIS-REx radio communication system (McMahon et al., 2018), the mission produces four maps of decision-making properties: deliverability; safety; sampleability; and scientific value (Lauretta et al., 2017). Each of these maps play a key role in selecting the OSIRIS-REx sample-site. The deliverability map illustrates the probability that the OSIRIS-REx flight dynamics team can deliver the spacecraft to a desired area on the surface of Bennu. The safety map estimates the probability that physical hazards are present at a location. The sampleability map quantifies the probability that a sample can be successfully collected from a given point on the surface. Finally, the scientific value map should contain as many key properties as possible to address questions and hypotheses posed by the sample analysis team within a geological context that is definitive enough to determine sample history.

For sampleability, safety, and scientific value decisions, imaging data provides spatial context and a quantitative assessment of the nature of the asteroid surface. Images of Bennu culminate in several foundational products, including photogrammetrically controlled and photometrically corrected mosaics or basemaps (Section 3.1-3.2), orthoimages, and digital terrain models (DTMs) (Section 3.3-3.4). Foundational products, and the data derived from them, will inform the downselection from global observations of Bennu to two candidate sample-sites: a prime and a backup. These are key in illustrating the geospatial distribution of hazards, potential safety concerns, and regions of interest across the asteroid. As a result, securing the data necessary to create these products is a high priority for the mission. Observational constraints and imaging campaigns are designed to acquire inputs that are optimized for particular data products. Due to the challenges of spacecraft navigation around a small body, achieving the desired observations has required close coordination between multiple elements of the mission.

Performing image-based mapping and generating cartographic products of a small body also poses challenges from a data processing perspective. At the inception of the OSIRIS-REx mission in 2011, the ability to perform accurate image processing of small bodies was largely unsupported by software packages used for planetary data analysis (e.g. SPICE, *ISIS3*, SOCET SET, JMARS, etc.). The OSIRIS-REx mission has provided an impetus for the development of tools and techniques capable of processing images from irregularly shaped bodies. We are hopeful that these efforts will benefit upcoming investigations such as those conducted by the Martian Moons eXploration (Kuramoto et al., 2017), Lucy (Levison, 2016), Psyche (Elkins-Tanton et al., 2017), and DESTINY+ (Arai et al., 2018) missions.

To evaluate the image processing methods and techniques needed to support site selection at asteroid Bennu, we adopted requirements definition tools from the discipline of systems engineering. The OSIRIS-REx mission Level-2 science requirements provide the basis for this framework, from which we specify the data products that are needed. On OSIRIS-REx, this was performed in tandem with the development of OCAMS (Rizk et al., 2018). This approach outlines the relationship between science goals, instrument design, data products, spacecraft observations, and ground data processing. It enables traceability for evaluating the strategies that are needed for imaging Bennu, planning campaigns to obtain that data, and implementing the techniques to perform image-based mapping of the asteroid through ground processing. It also provides a disciplined framework for making trades, and has allowed us to balance a desire to acquire the best data possible with the demands of quickly and safely sampling. We describe the advancements made by applying this approach to the challenges associated with image-based mapping of small planetary bodies in preparation for the OSIRIS-REx mission. In particular, we focus on observational strategies and new image processing techniques that we have established to support the overarching goal of creating products that will aid a site-selection decision.

## 2 OCAMS Overview

Scientific images of Bennu are acquired with OCAMS, a trio of cameras that are optimized for encounter with asteroid Bennu. The capabilities of OCAMS are comprehensively detailed in Rizk et al. (2018). During asteroid proximity operations, OCAMS images are critical in identifying candidate sampling sites and documenting the Touch-and-Go (TAG) sampling event. Collectively the low- (SamCam), medium- (MapCam), and high-resolution (PolyCam) imagers provide long-range acquisition of Bennu as a point source, perform the global mapping and shape determination, characterize the sample-site at sub-centimeter scales, and, finally, record the sampling event and verify post-sampling success.

**Table 1.** Summary of the MapCam and PolyCam Imagers. Properties of PolyCam are given at the two ends of its focal range, 200 m and infinity.

|  | **MapCam** | **PolyCam (200 m)** | **PolyCam (∞)** |
|---|---|---|---|
| Camera optical design | 5-element refractor | variable focus Ritchey-Chretien | |
| Detector type | 1024 × 1024 CCD | 1024 × 1024 CCD | |
| Pixel size (mm) | 0.0085 | 0.0085 | |
| Focal length (mm) | 125 | 610.2 | 628.9 |
| Field of view (mrad) | 69.6 | 14.3 | 13.8 |
| Aperture diameter (mm) | 38 | 175 | |
| Pixel size 1 km from surface (cm) | 6.8 | 1.4 | |
| Broadband filter range (nm) | 489-817 | 489-808 | |
| Effective $b'$ filter wavelength (nm) | 473 | - | - |
| Effective $v$ filter wavelength (nm) | 550 | - | - |
| Effective $w$ filter wavelength (nm) | 698 | - | - |
| Effective $x$ filter wavelength (nm) | 847 | - | - |

Each camera is designed with a specialized function to support the sample return objective. Image-based mapping of Bennu is almost exclusively performed using MapCam and PolyCam (Table 1), which have fields of view (FOV) that differ by a factor of five, thereby allowing interspersed MapCam images to provide context data for higher resolution PolyCam frames in the same observation. MapCam also provides color data of Bennu using four narrow-band filters and panchromatic data using a broadband filter. The narrow-band filters closely approximate the Eight Color Asteroid Survey (ECAS; Zellner et al., 1985) *b-, v-, w-*, and *x*-bands (Table 1). The exception is the MapCam *b'* filter, which has a longer effective wavelength than its ECAS counterpart, but is better suited to the optical and radiometric performance of MapCam. PolyCam is the primary tool for high-resolution panchromatic mapping of Bennu's surface. It has a variable focus that enables it to capture Bennu from a distance of several million kilometers and can also image the asteroid at sub-centimeter scale from ranges of 200 m. Finally, SamCam documents the interaction of sample-site and the Touch-and-Go Sample Acquisition Mechanism (TAGSAM) before, during, and after the TAG maneuver and provides backup context imaging capabilities.

Within the ground data processing system, OCAMS data are calibrated in units of radiance, spectral radiance, and reflectance (also known as radiance factor or I/F). Images in units of reflectance are the primary input to the products described here.

## 3 OCAMS Data Products of Asteroid Bennu

Basemaps, orthoimages, and DTMs are considered foundational data products. Their creation is recommended for planetary missions by the most recent Planetary Science Decadal Survey (Space Sciences Board, 2012) and advocated for by a number of advisory groups (e.g. Archinal et al., 2011; Lawrence et al., 2015). Foundational products have been instrumental in the success of prior planetary missions, and particularly landed missions (e.g., Grant et al., 2011; Kirk et al., 2008; Spencer et al., 2009). For the OSIRIS-REx mission, these products serve as the underpinnings of higher-order scientific investigations of Bennu, including geological mapping and studies of surface processes. Moreover, they provide operational utility to the mission by furnishing properties of the asteroid that clarify engineering constraints. As such, these products can be directly linked to the Level-2 science requirements (Table 2). Here we provide an overview of these products as they are generated throughout the mission and emphasize their importance in the sample-site decision. Image-based products such as photometric models and shape models derived via stereophotoclinometry also play a major scientific and operational role in the success of the mission and are well detailed in our companion paper by Golish et al. (2018) and in Weirich et al. (2016), respectively.

### 3.1 Panchromatic Basemaps

At both global and site-specific scales, panchromatic (broadband filter) basemaps illustrate the highest-resolution data of Bennu available from the OSIRIS-REx instrument payload. As a result, they are the starting point for several thematic maps and long-term science products. Panchromatic basemaps supply the high-resolution data needed to assess the sampleability of regolith across the surface of Bennu. They also provide a basis for geologic and geomorphic maps of the asteroid and context for lower-resolution spectral measurements. These products are created primarily with PolyCam images that have been photogrammetrically controlled (Section 5.3). They are refined throughout the mission as higher-resolution data are accumulated leading up to the sampling event.

The global basemap of Bennu is generated from PolyCam images that achieve < 5.25 cm per pixel, resulting in a basemap spatial resolution < 21 cm (using a 4-pixel criteria). The spatial resolution requirement is driven by the need to understand the distribution of hazardous surface features ≥ 21 cm in longest dimension (in particular boulders and blocks). Here we define a hazard as an object ≥ 21 cm or any other feature that might pose a threat to the integrity of the TAGSAM (Bierhaus et al., 2018). Specifically, 21 cm represents the minimum dimension that can obstruct the inner annulus of the TAGSAM head (Figure 1) thereby preventing the acquisition of a sample that can fit safely into the OSIRIS-REx Sample Return Capsule (SRC). Aside from blocks and boulders, examples of hazards may include craters, grooves, or linear features that possess significant surface slopes; the maximum surface angle that TAGSAM can mechanically accommodate is 14°. As a result, it is imperative that features ≥ 21 cm and other potential hazards be precisely characterized within regions of interest (areas of relatively smooth terrain or high scientific value) ahead of the site selection decision.

**Table 2.** The relationship between the OSIRIS-REx mission Level 2 requirements, spacecraft observations, and image-based mapping products. Level 2 requirements are abbreviated for clarity. Approximate altitudes of imaging campaigns are defined relative to the body-center, except for the Reconnaissance campaign, where they are defined relative to the body-surface.

| Driving OSIRIS-REx requirement | Imaging campaign and altitude | Primary camera | Local solar time (solar longitude) | Pixel size (cm) | Data products generated |
|---|---|---|---|---|---|
| OSIRIS-REx shall image >80% of the surface of Bennu with <21cm spatial resolution to assess the presence of hazards and regions of interest. | *Baseball Diamond:* Panchromatic Strategy, 3.7 km | PolyCam | 10:00 AM (-30°) | 5 | Global panchromatic basemaps; global survey of hazards |
| OSIRIS-REx shall, for >80% of the asteroid surface, map the blue slope, visible slope, and the depth of the 700 nm absorption feature at <1 m spatial resolution. | *Baseball Diamond:* Color Strategy, 3.8 km | MapCam | 12:30 PM (7.5°) | 24 | Global color basemaps |
| OSIRIS-REx shall, for >80% of the asteroid surface, map the variation in spectral properties in regions where the albedo is >1% using photometrically corrected (to 30° phase angle) and normalized MapCam reflectance spectra, for each filter from 470-850 nm. | *Equatorial Stations,* 5 km | MapCam | 6 AM (-90°), 10 AM (-30°), 12:30 PM (7.5°), 3 PM (45°), 6 PM (90°), 8:40 PM (130°), 3:20 AM (-130°) | 35 | MapCam photo-metric models in the *b', v, w, x* and panchromatic filters. |
| OSIRIS-REx shall select a sample-site that satisfies the following: 99% probability of ensuring the safety of the flight system during sampling; ≥80% probability of acquiring ≥60 g of bulk sample per sampling attempt. | *Baseball Diamond:* Color Strategy, 3.8 km | MapCam | 12:30 PM (7.5°) | 24 | *x*-band mosaic for 1064 nm basemap |
|  | *Orbital B,* 1 km | PolyCam | 6 PM (90°), 6 AM (-90°) | 1 | Context images to assess ≥ 5 cm PSFD |
| OSIRIS-REx shall, for >80% of a 2σ TAG ellipse around at least the prime sampling site, map the blue slope, visible slope, and the depth of the 700 nm absorption feature at <25 cm spatial resolution. | *Reconnaissance:* 525 m Flyby | MapCam | 12:30 PM (7.5°) | 3.5 | Site-specific color basemaps |
| OSIRIS-REx shall, for >80% of a 2σ TAG ellipse around at least 2 candidate sampling sites map the areal distribution and determine the particle size-frequency distribution of regolith grains ≥ 2 cm in longest dimension. | *Reconnaissance:* 225 m Flyby | PolyCam | 2:30 to 3:30 PM (37.5° to 52.5°) | 0.3-0.5 | Site-specific panchromatic basemaps; PSFDs |
|  |  | MapCam | 2:30 to 3:30 PM (37.5° to 52.5°) | 1.5 | Stereo images, DTMs, orthoimages |

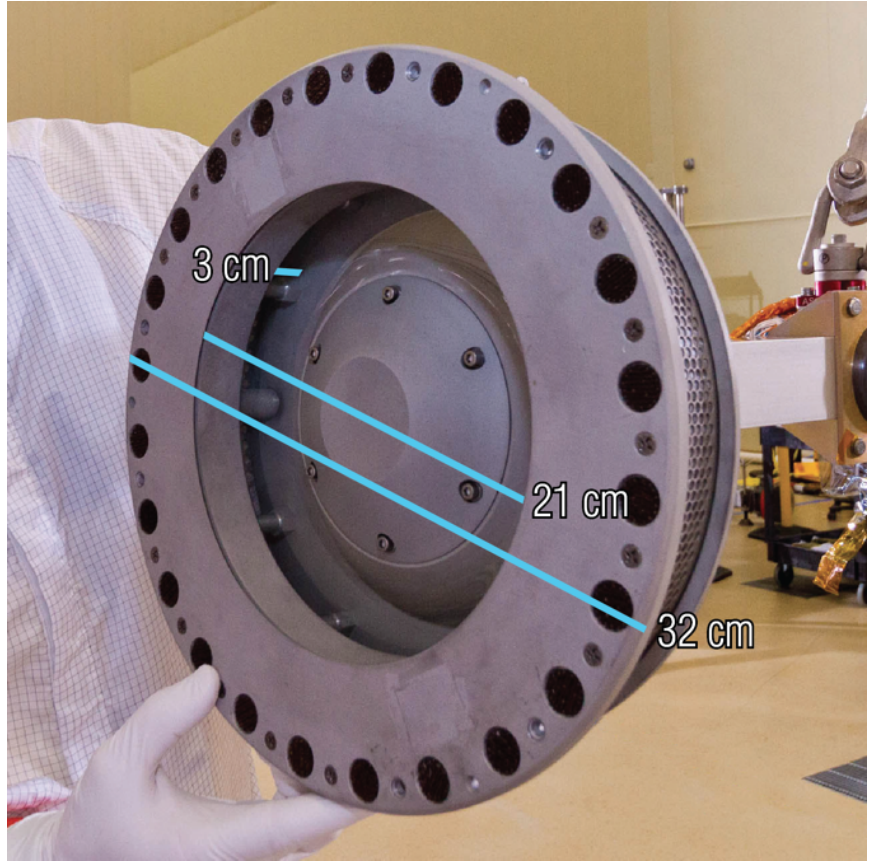

**Figure 1.** The dimensions of the TAGSAM head, which prescribes the resolution requirements for the global and site-specific imaging campaigns of Bennu. Objects ≥ 21 cm can obstruct the inner annulus of the TAGSAM head, and grains ≤ 2 cm are readily ingested by TAGSAM through its annular throat.

The next set of panchromatic basemaps uses PolyCam images at < 2 cm spatial resolution (< 0.5 cm/pixel) from site-specific 225 m flybys. The purpose of these site-specific basemaps is to illustrate the distribution of regolith grains 2 cm in longest dimension at the two candidate sample-sites and detect any features ≥ 5 cm in longest dimension. Grains ≤ 2 cm are smaller than the narrowest section of the annular throat of the TAGSAM (Figure 1), and hence represent "sampleable" material that can be readily ingested during the TAG maneuver. The maximum height obstruction with which the TAGSAM head can interface and not exceed its tilt budget is 5 cm. Using the site-specific panchromatic basemaps, particles are counted, measured, and georeferenced down to the 4-pixel limit of the images. This provides the regolith particle size frequency distribution (PSFD) at the prime and backup sample-sites, which aids in predicting the relative sampleability of each.

### 3.2 Color Basemaps

A series of color basemaps are produced using MapCam's four color filters (Table 1). As with the panchromatic basemaps, data for color basemaps are acquired at both global and site-specific scales. Despite the limited spectral information, MapCam color images provide the highest spatial resolution of any spectral data set obtained by the OSIRIS-REx payload. Color images are an established tool for elucidating surface composition (e.g., Delamere et al., 2010;

Le Corre et al., 2013), regolith grain size (e.g., Clark et al., 2010; Jaumann et al., 2016), and space weathering signatures (e.g., Chapman et al., 2004; Hiroi et al., 2006), and hence they help assess the relative scientific value across Bennu. However, much like the panchromatic data, color images also serve an operational purpose as the spacecraft team leverages them to evaluate the safety of Bennu's surface. In particular, *x*-band frames are used to examine the surface at longer optical wavelengths to sense bright terrains that might overwhelm the Guidance Navigation and Control (GNC) flash lidars during TAG. At present, three different types of basemaps are created using MapCam color filter data: color-index maps, true-color maps, and an *x*-band mosaic extrapolated into a 1064 nm basemap.

Color-index basemaps are mosaics that highlight spectral variations across the surface of Bennu at significantly finer spatial scales than OVIRS. These include maps created with a single color parameter such as a ratio of two filters (e.g., Figure 2a), and false-color composite maps that associate a specific color parameter with a color channel to form an RGB image. A preliminary set of color-indices have been identified to study Bennu's surface composition and space weathering effects using MapCam's four color filters. These indices include *b'/v* to characterize the ultraviolet to visible (blue) slope, *v/x* to characterize the visible to infrared (visible) slope, the turn-off point (wavelength of the maximum point on 2nd order polynomial fit to the four filters), and an expression to characterize the presence and depth of the 700 nm absorption feature, which is indicative of phyllosilicates (Vilas, 1994). Specifically, we assess the presence of the 700 nm absorption feature by determining the relative band depth (RBD) using an expression adopted from Bell and Crisp (1993):

$$RBD = 1 - \left\{ \frac{\frac{I_{\lambda w}}{\bar{I}_{\lambda w}}}{(1-f)\left(\frac{I_{\lambda v}}{\bar{I}_{\lambda v}}\right) + f\left(\frac{I_{\lambda x}}{\bar{I}_{\lambda x}}\right)} \right\} \qquad (1)$$

Where $I$ represents the reflectance of wavelengths $\lambda_v < \lambda_w < \lambda_x$, $\bar{I}$ is the global average reflectance value at each wavelength, and $f = \frac{\lambda_w - \lambda_v}{\lambda_x - \lambda_v}$. The denominator represents the local continuum at $\lambda_w$ and is derived from the linear fit of the *v*- and *x*-band reflectance obtained on either side of the 700 nm absorption feature.

We create false-color basemaps by assigning the red slope (*v/b'*) to the red channel, the 700 nm band depth to the green channel, and the blue slope (*b'/v*) to the blue channel of an RGB composite image. These RGB composites are similar to those created by the Clementine (e.g., Pieters et al., 1994) and Dawn (e.g. Reddy et al., 2012) missions, in which the redder regolith appears red and the material with a bluer slope appears blue, while the green channel is indicative of surface composition. Comparisons between MapCam color units and the spectral indices measured by OVIRS are also conducted to verify compositional trends. Other color-indices and RGB combinations that are deemed diagnostic, especially those related to signatures of space weathering, may also be used to investigate the distribution of color units on Bennu. Principal Component Analysis (PCA), a method applied previously to MESSENGER multi-spectral images of Mercury (Robinson et al., 2008), is used to search for spectral variations in the 4-band color data. PCA can group terrains with similar spectral properties together and help discriminate subtle spectral differences in homogeneous regions of Bennu's surface. Spectral clustering techniques will also be applied to achieve a similar end (e.g., Calinski and Harabasz, 1974).

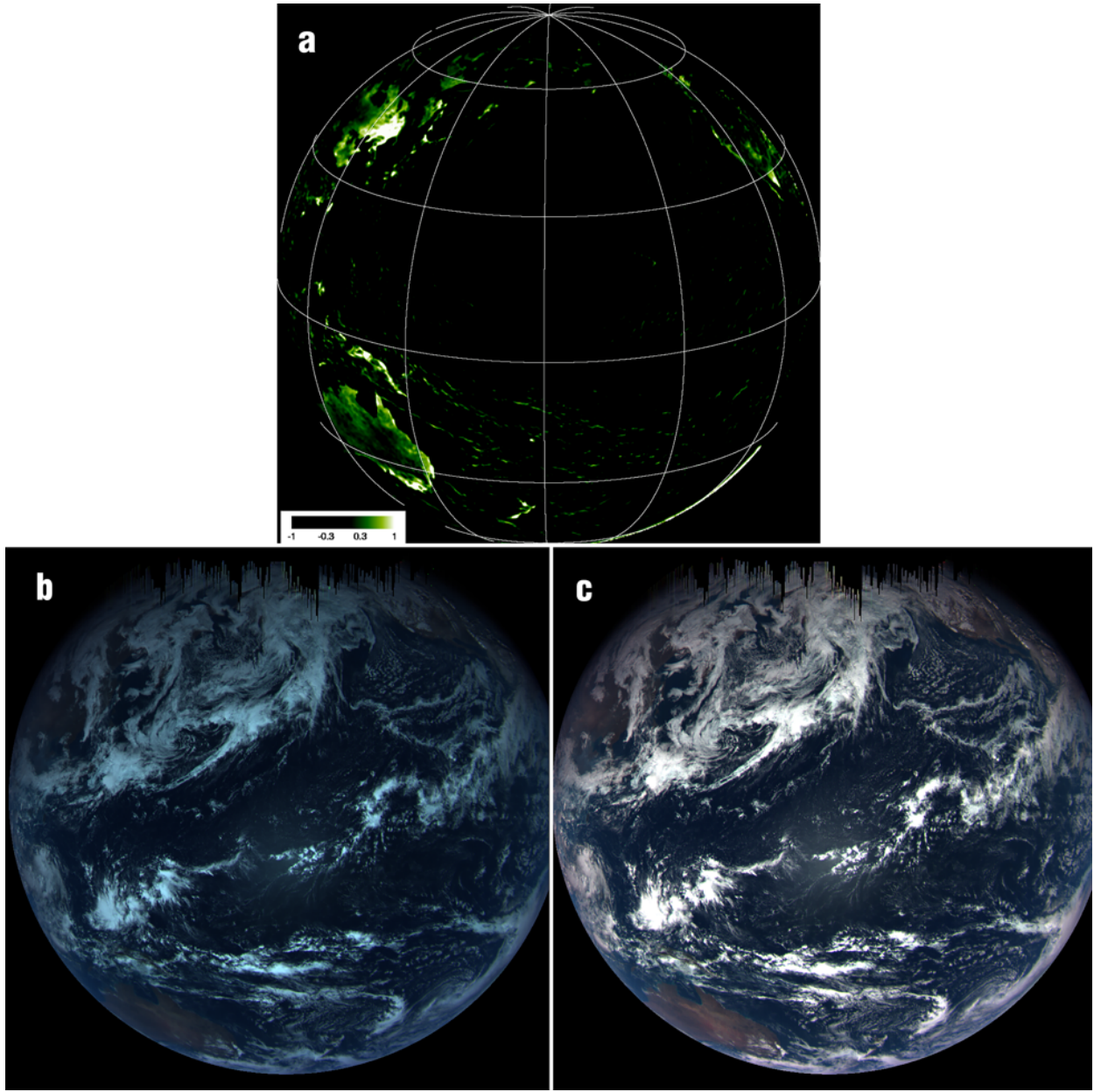

**Figure 2.** Examples of MapCam color composites from the OSIRIS-REx Earth Gravity Assist. (a) A normalized difference vegetation index composite derived from the MapCam *w* (698 nm) and *x* (847 nm) band images. The white to bright green areas indicate a higher density of vegetation compared to darker green areas. (b) An RGB composite of *w, v,* and *b'* prior to color correction. The *w* (698 nm), *v* (550 nm), and *b'* (473 nm) filters are assigned to the R, G, and B channels respectively (c) An RGB composite of *w, v,* and *b'* corrected to provide a more natural color rendition by performing white balancing on clouds within the image frames; that is, by scaling the green (*v*) and red (*w*) filter images with respect to the blue (*b'*) image until the clouds visually appear white.

      True-color basemaps illustrate Bennu's surface in a natural color rendition by assigning the *w* (698 nm), *v* (550 nm), and *b'* (473 nm) filters to the red (R), green (G), and blue (B) channels, respectively. To create true color images of the Earth obtained during the OSIRIS-REx

Earth Gravity Assist, images were color balanced using a set of correction factors obtained by white balancing MapCam cloud data (Figure 2b, 2c). This correction is necessary given that MapCam filters do not perfectly align with typical RGB models. The same correction factors will be applied to Bennu to create true-color composites.

To aid the production of the safety map, *x*-band mosaics are used as a foundation for the 1064 nm basemap. During TAG the GNC lidars autonomously guide the spacecraft to the surface for sample acquisition. The GNC lidars measure the surface bidirectional reflectance at 1064 nm at a 0° phase angle (i.e., the incident and reflected rays are parallel). To evaluate the relative safety of terrains on Bennu it is necessary to predict the surface reflectance at 1064 nm and identify areas where reflectance exceeds the operational tolerances of the lidars. Predictions of the spatial distribution of intensity are also important for avoiding areas with high variation that could affect lidar ranging performance. To help with these assessments, low-phase, *x*-band images are photometrically corrected, converted from I/F to a Bidirectional Reflectance Distribution Function (BRDF), and mosaicked to provide a basemap of Bennu's surface at the longest optical wavelength that can be sensed by OCAMS. This *x*-band (847 nm) mosaic is translated to a 1064 nm basemap by combining these data with lower spatial-resolution OVIRS spectroscopic measurements of the 860-to-1064 nm ratio across the asteroid.

### 3.3 Orthoimages

Along with basemaps, orthoimages are generated from nearly all surface-resolved OCAMS images of Bennu. Orthoimages are frames that are geometrically corrected (orthographically rectified) to a uniform scale. Orthorectification uses terrain measurements (a shape model or a local DTM) to reproject images as though they were taken under nadir viewing conditions, thereby removing any geometric effects from topography, oblique camera angles, or lens distortion. The uniform scale allows orthoimages to be used as map products so long as body-fixed coordinates are associated with each pixel.

High precision sample-site orthoimages are created from MapCam stereo images for the two candidate sites using SOCET SET, an established commercial software platform from BAE Systems, Inc. These images are orthorectified using the terrain model generated from their corresponding stereo strength (Section 3.4). Lower precision orthoimages are produced throughout the mission using OCAMS camera models and a 3D tessellated shape model of Bennu. To support this functionality, we enhanced the Integrated Software for Imagers and Spectrometers version 3 (*ISIS3*; Keszthelyi et al., 2013) developed by the U.S. Geological Survey to improve the orthorectification for images from small bodies (Section 5.2). These orthoimages are map projected and serve as inputs to larger mosaics and basemaps. They are also converted into image stamps for the OSIRIS-REx data visualization tool, J-Asteroid, which is based on the Java Mission-planning and Analysis for Remote Sensing (JMARS) software platform (Hagee et al., 2015).

### 3.4 Digital Terrain Models from Stereophotogrammetry

For much of the OSIRIS-REx mission, DTMs and global asteroid shape models are generated from OLA ranging data, stereophotoclinometry (SPC; Gaskell et al., 2008) of OCAMS images, and combinations of these techniques. DTMs developed through stereophotogrammetry (SPG) are intended to provide a backup to the models derived through SPC. Use of independent techniques is expected to provide a robust description of the terrain, and is particularly important for evaluating potential sample-sites. Consequently, DTMs from SPG are only needed to examine the prime and backup candidate sites and are derived from site-specific MapCam stereo images with an expected ground sample distance (GSD) of ~6 cm. Lower GSD DTMs (~20 cm) from global PolyCam stereo data may also be derived to complement the results from SPC.

DTMs for candidate sample-sites are generated using SOCET SET and *ISIS3*, following procedures adapted from those developed for the MESSENGER Mercury Dual Imaging System (MDIS) framing camera (Hawkins et al., 2007; Manheim et al., 2017). SPG starts with OCAMS image pairs that meet well-defined stereo and illumination criteria (e.g., Becker et al., 2015). Targeted stereo pairs should have a convergence angle of 10º–30º; their incidence angles should be between 20º and 70º and not differ by more than 10º. A smaller convergence angle is best for very rugged terrain, to avoid obscuring parts of the surface. A larger convergence angle is optimal for very flat, smooth terrain; the larger parallax makes estimation of elevation values more robust. The incidence angle range optimizes illumination with shadows. On an airless body, such as Bennu, deeply shadowed areas will not be illuminated from light reflecting off an atmosphere, so it is desirable to minimize deep shadows in stereo pairs. Image resolution is another criterion for stereo pair selection. Feature matching algorithms used in stereo correlation will produce more accurate alignment for images with similar resolutions, not different by more than a factor of two.

The stereo pair is preprocessed in *ISIS3* to generate the input for SOCET SET, including the spacecraft position and camera pointing angles. The images are first controlled relative to each other in order to produce an initial terrain model. The resulting model is then referenced to OLA measurements to provide absolute vertical control using an auto-triangulation algorithm that produces a best-fit solution between the DTM and the altimetry (Kilgallon et al., 2015). We maintain the ability to solve for terrain height using a local rectangular coordinate system that is independent of a datum. Even so, sample-site DTMs generated via SPG represent relatively small, local-scale patches of ground that are expected to be smooth and free of hazards, and can likely be expressed relative to a sphere at the local radius, thereby allowing us to retrieve local elevation. DTMs will be produced at a GSD of four times the nominal pixel scale of the source stereo pair. For example, a stereo pair acquired at a pixel scale of 5 cm will result in a DTM with a GSD of 20 cm. The expected vertical precision (EP) of the elevation values can be estimated based on the convergence angle, spatial resolution of the stereo pair, and the quality of feature matches (Kirk et al., 2008; Sutton et al., 2015). Given optimal stereo pair illumination and geometry, the EP of OCAMS DTMs will be on the order of a few centimeters for a DTM with a GSD of 20 cm. Testing with simulated data shows, for example, that for a stereo pair with a convergence of 20º, the EP is ~1–2 cm. SOCET SET provides a measure of the relative linear error (the expected error from one elevation point to any other with 90% confidence) in a DTM (BAE Systems, 2009; Henriksen et al., 2017). For the simulated data tests, the relative linear error is ~10 cm for a 20-cm DTM. The horizontal and vertical accuracy will be on the order of that of the OLA data, which will provide the absolute control for the DTM. Primary sources of topographic errors can include bland, deeply shadowed, or saturated areas which can generate stereo matcher interpolation artifacts. The quality of stereo matches as well as interpolated or manually edited areas are identified in the SOCET SET Figure of Merit (FOM) maps (Mattson et al., 2012). Stereo pairs with poor geometry (e.g., low convergence angle for flat terrains) or large offsets in illumination can also complicate SPG processing and terrain quality. After quality assessment and any manual editing, the source stereo images are orthorectified to the DTM at the full spatial resolution of the source images. The DTM and orthoimages are exported as raster files and post-processed in *ISIS3* to apply map definitions consistent with the team requirements for sample-site analysis.

## 4 Observational Strategies for Mapping Asteroid Bennu

### 4.1 Overview

Several navigational complexities must be overcome to ensure that the OSIRIS-REx spacecraft can rendezvous with and acquire data of asteroid Bennu. Spanning only ~500 m in the longest dimension (Nolan et al., 2013), Bennu has the lowest gravitational acceleration of any planetary object a spacecraft has attempted to orbit (GM of $5.2 \pm 0.6$ m$^3$/s$^2$, (1σ); Chesley et al., 2014). In this low gravity environment, the acceleration due to solar radiation pressure significantly influences the dynamics of the spacecraft, which tenuously orbits Bennu with relative velocities < 10 cm/s (Wibben et al., 2017). To achieve stability against solar radiation pressure, quasi-periodic orbits around Bennu must reside in the terminator plane. Imaging from a terminator orbit, however, negatively impacts data quality. Images acquired at a nadir-pointed geometry (ideal for minimizing projection effects) are at or near the terminator and hence poorly illuminated. Under these conditions, terrain will cast long shadows that can obscure surface features and the signal-to-noise ratio (SNR) of images is likely to be lower. Conversely, images acquired when the camera is pointed away from the terminator (at larger emission angles) will achieve more favorable illumination but suffer from effects such as foreshortening, reduced image resolution, and a loss of focus if the terrain extends outside the camera's depth of field. To avoid these data-acquisition compromises, OCAMS primarily images Bennu's surface through a series of hyperbolic flybys that have significantly more favorable viewing geometries than orbit. Many of these flybys are designed to acquire data that are optimized for mapping at global and at site-specific scales. In this section we describe the imaging campaigns planned throughout the Science Phase of the mission.

### 4.2 Concept of Operations for Mapping Bennu with OCAMS

The initial images of Bennu are acquired during a tightly scripted Navigation Phase of the mission, which takes place in three parts: Approach, Preliminary Survey, and Orbital A (Lauretta et al., 2017). Each part of the Navigation Phase is intended to provide increasing knowledge of the asteroid's ephemeris, gravity field, and shape to support the transition from star-based to landmark-based optical navigation (Wibben et al., 2017). The Navigation Phase of the mission concludes near the end of Orbital A when the Science Phase is initiated.

In the Science Phase, we collect the OCAMS images that are needed to construct data products derived from the mission's Level-2 science requirements (Table 2). Consequently, strategies have been identified to collect observations that are optimized for particular products. The first portion of the Science Phase is Detailed Survey, which is divided into two campaigns: Baseball Diamond and Equatorial Stations. Baseball Diamond comprises seven hyperbolic flybys that are designed to acquire images for either basemaps or SPC (the Baseball Diamond name is historical, as the campaign once consisted of four equidistant observing stations). Equatorial Stations includes seven flybys intended for spectral mapping with the OTES and OVIRS and for photometric analysis in each MapCam filter. To enable photometric modeling, Equatorial Stations acquires resolved imaging across a range of phase angles as is well-detailed by Golish et al. (2018).

The seven Baseball Diamond flybys fulfill three separate observational strategies (Figure 3). Two of these strategies provide inputs for the color and panchromatic basemaps of Bennu using MapCam and PolyCam, respectively. The third provides low-phase context data for SPC. The MapCam color imaging strategy includes two 3.8 km flybys comprised of three stations, which observe the asteroid at 12:30 PM local time (7.5° solar longitude) (see Figure 3b). These three stations acquire low-emission and low-phase-angle observations over Bennu's 4.3-hour rotation period. Bennu is imaged in each of MapCam's five filters resulting in a global color data set of the asteroid's surface with SNR > 100. The 12:30 PM local solar time is chosen to provide low incidence and phase angles that highlight Bennu's albedo variations in each color filter, while muting the appearance of shadows and terrain. The 30 minute local solar time offset (7.5°) from 12 PM ensures that images do not view the opposition effect even when accounting for navigational uncertainties. Image sets in all five filters are acquired in rapid succession to capture the same scene under near-uniform photometric conditions, thereby maintaining the spectral integrity of color cubes. In subsequent ground data processing the four-colors are co-registered with a panchromatic frame and combined into a multispectral cube with < 1 m spatial resolution (< 25 cm/pixel). Compared to other data-acquisition phases, the spacecraft slowly slews during the MapCam color-imaging strategy to provide as much filter-to-filter image overlap as possible. Even when the spacecraft is perfectly stationary relative to the asteroid, Bennu's rotation offsets successive MapCam-filter frames by approximately four pixels.

PolyCam panchromatic imaging during Baseball Diamond is similar to the color-imaging strategy but with the addition of two flybys that provide stereo (Figure 3c). Initially, two 3.7 km flybys comprised of three stations each are conducted at 10 AM local time (-30° solar longitude). These three stations acquire low-emission and mid-phase-angle observations over Bennu's 4.3-hour rotation period for the north, south, and equatorial latitudes resulting in a global base map of Bennu at < 21 cm spatial resolution (< 5.25 cm/pixel). Imaging at a local time of 10 AM results in images with incidence angles between 30-60° at equatorial through mid-latitudes. These midrange incidence angles provide data that highlight terrain and illustrate the relief across the surface, which can be used to define the characteristic shape of features such as boulders and craters. Stereo complements to these data are obtained at the same local solar time (solar longitude), but at a stereo angle of approximately 20°, during two subsequent flybys that cover the northern and southern hemispheres. Stereo data help constrain the shape of the asteroid and improve the image network geometry (Section 5.3). Individual stereo pairs are also used to produce DTMs of regions of interest and provide backup capabilities to those provided by SPC. Our plan for stereo coverage follows the recommendations of Schaller and Milazzo (2009) and Becker et al. (2015). Because of the narrow field of view of PolyCam, slewing is performed to obtain north-south coverage at each imaging station. Exposure times are chosen to limit motion blur to less than one pixel while achieving an SNR of ~100. During the equatorial flyby, panchromatic MapCam data are also obtained at 10 AM local solar time to provide context for the smaller PolyCam frames.

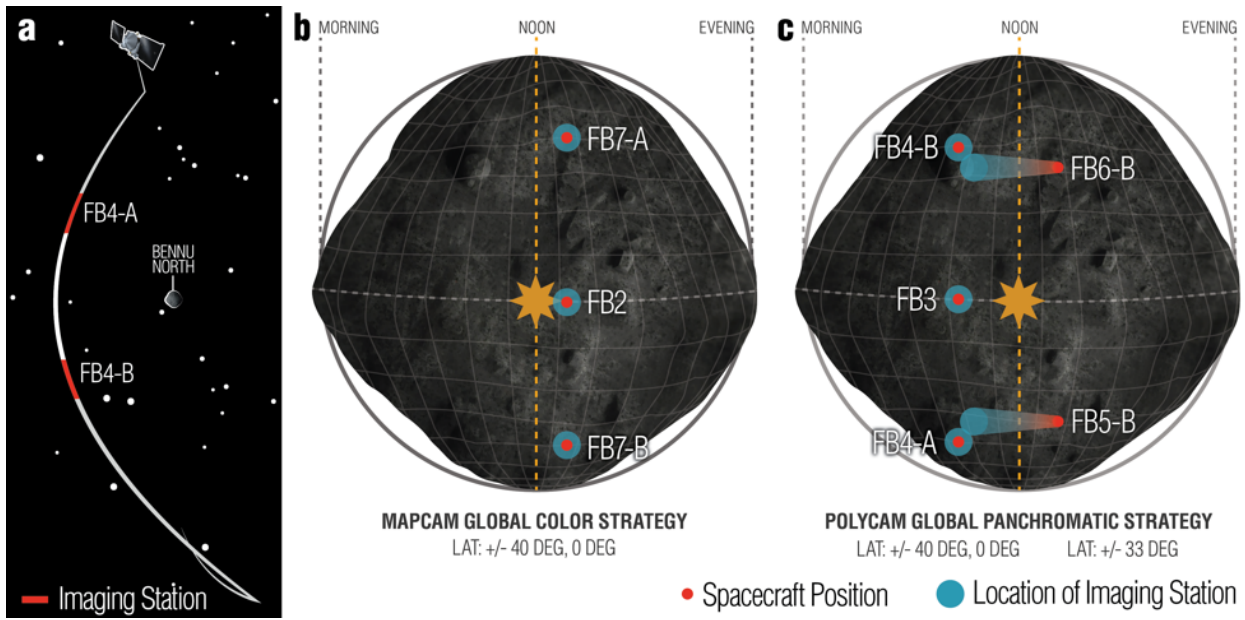

**Figure 3.** (a) Schematic of the Detailed Survey Baseball Diamond hyperbolic trajectories that target a series of imaging stations (red) within a flyby. (b, c) Cartoons depicting a subset of the Baseball Diamond imaging stations that observe the surface of Bennu from a specific latitude and local solar time. Each station is designated by the flyby number ('FB#') and the order in which they are obtained during that flyby ('-A' or '-B'). (b) Illustrates the three stations that acquire global color images during two flybys ('FB2' and 'FB7'). (c) Illustrates the panchromatic imaging strategies, which is similar to the color-imaging strategy, but incorporates two additional stations ('FB6-B' and 'FB5-B') that provide stereo coverage.

At the end of Detailed Survey the spacecraft transitions to a 1 km terminator orbit, designated Orbital B (Figure 4). During this campaign the spacecraft acquires global OLA data and gravity field measurements. Due to the limited utility of images taken from the terminator plane, OCAMS data for map products are not collected during Orbit B. However, sparse, globally distributed context images from PolyCam are acquired to estimate the size frequency distribution of boulders and cobbles at the ~5 cm scale. Using this data, we also determine the correlation, if any, between the size frequency distribution and the roughness measured by OLA at a ~7 cm baseline. These data aid relative assessments of sampleability before the selection of two candidate sample-sites is made.

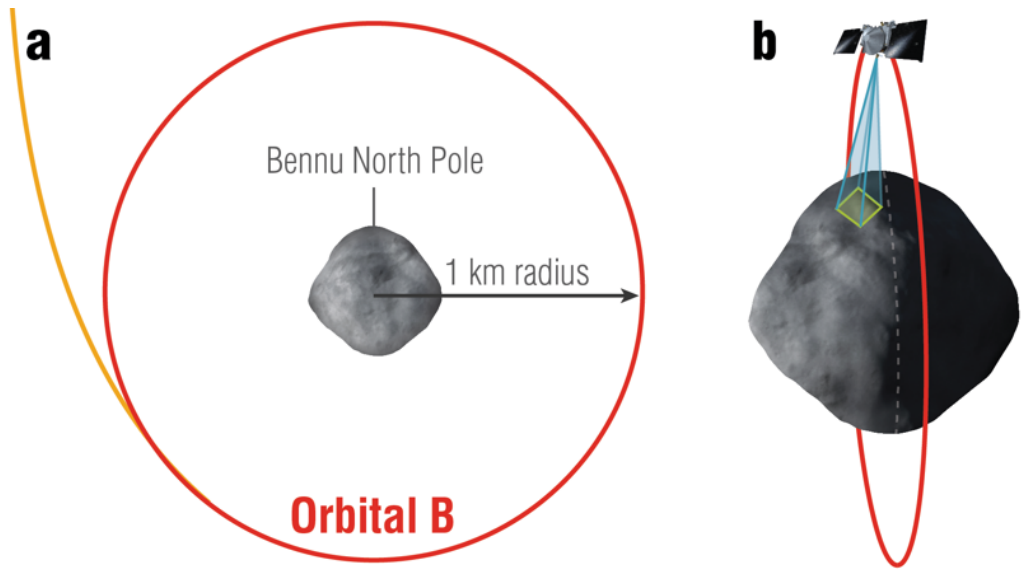

**Figure 4.** (a) Schematic illustrating the Orbital B 1 km terminator orbit. (b) Observations from Orbital B acquire globally distributed high-phase angle PolyCam images to help assess the population of ~5 cm regolith grains ahead of the downselection to two candidate sample-sites.

The final segment of the Science Phase is Reconnaissance. Reconnaissance consists of several low-altitude flybys that characterize the candidate sample-sites chosen at the end of Orbital B. During this phase, 525 m range-to-surface flybys observe both candidate sites in each MapCam filter, resulting in local color maps at 14 cm spatial resolution (3.5 cm/pixel). Two additional sets of 225 m flybys acquire panchromatic PolyCam images of each site (Figure 5). Due to the close range of the 225 m flyby and the size of the PolyCam FOV, the spacecraft rapidly slews to achieve 80% coverage of a 2σ ellipse around both sites (based on the navigational uncertainty). As observations from a single flyby may provide incomplete (<80%) coverage, a second 225 m flyby of each site is necessary to meet requirements. MapCam stereo pairs are also acquired at the 225 m range to provide inputs to SPG and compensate for coverage gaps. Panchromatic basemaps at ~1.6 cm spatial resolution (0.3-0.5 cm/pixel) are generated from the images collected during each of the 225 m flybys, and their data are used to assess the population of regolith grains ≥ 2 cm. The particle size frequency distribution and sampleability of each site is a critical factor in selection of the final sample-site.

At the conclusion of the Science Phase, the primary sample-site is selected and the Sampling Phase of the mission is initiated. During the Sampling Phase, a series of rehearsals are conducted in preparation for the TAG maneuver. Throughout these events, OCAMS collects increasingly higher resolution context images of the prime sample-site, but no further mapping observations are targeted.

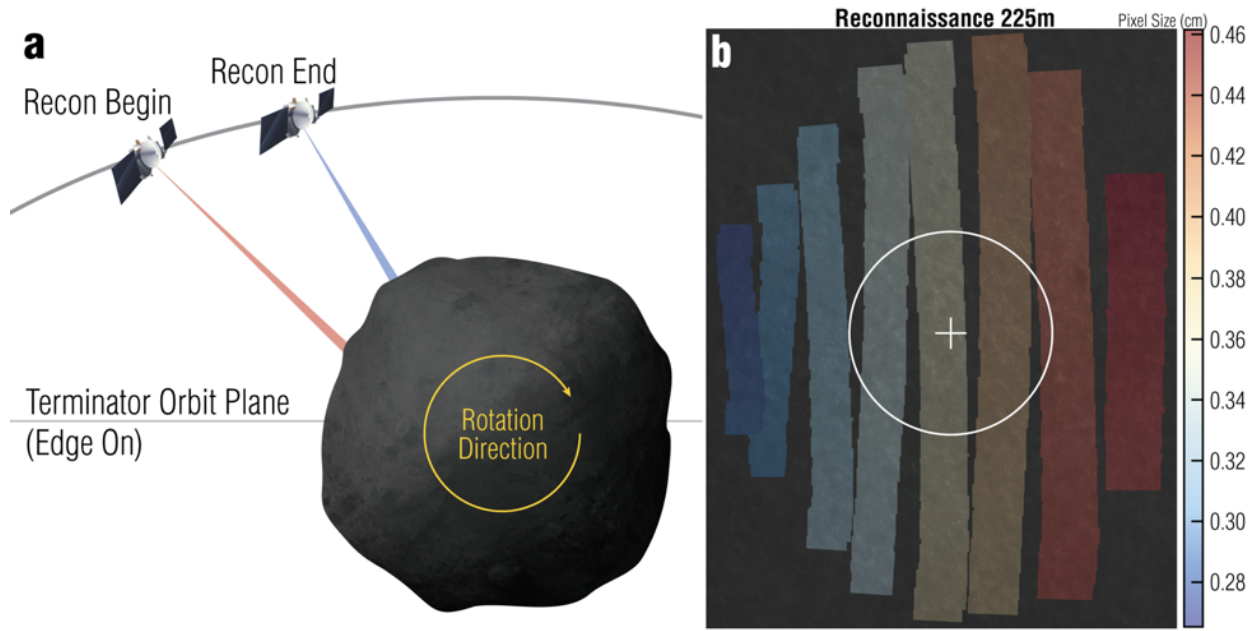

**Figure 5.** (a) Schematic illustrating the geometry of a 225 m Reconnaissance flyby over a candidate sampling site and (b) the resulting PolyCam image pattern, which illustrates the coverage obtained in a single flyby. Here, the white crosshatch indicates the targeted sampling site, and surrounding circle indicates the 2σ TAG ellipse. Multiple flybys are conducted to achieve thorough coverage of the sampling site across a TAG ellipse.

## 5 New Image Processing Techniques for Mapping Small Bodies

### 5.1 Overview

After OCAMS observations are downlinked, map products are created and visualized within the OSIRIS-REx mission ground data processing system. To conduct this work, we use several new processing techniques for mapping small bodies.

A planetocentric or planetographic coordinate system is suitable for planets or satellites that are near-spherical (or at least sufficiently convex) in shape (Nefian et al., 2013). Here, longitude and latitude coordinates uniquely define surface feature locations. In contrast, irregularly shaped bodies may have regions where surface features cannot be uniquely defined by longitude and latitude. Attempting to fit data from such objects to a spherical or ellipsoidal datum will result in data loss, visual distortion, and errors in image registration. In short, software assuming planetocentric or planetographic coordinates is inadequate for processing data from irregularly shaped bodies. A Cartesian coordinate system, in which points are expressed in rectilinear *x, y, z* coordinates is better suited to uniquely define feature locations.

Although Bennu's shape as determined through multiple radar apparitions is well understood down to the 20 m scale (Nolan et al., 2013), there are uncertainties in the polar radius that may result in Bennu appearing more flattened or elongated than current estimates dictate. At least one large topographic feature (~20 m) is apparent in the radar data (Nolan et al., 2013) indicating that some areas of overhanging terrain may exist on the asteroid. To prepare for the possibility that Bennu's shape significantly deviates from an ellipsoid, we have developed a set of planetary mapping tools that are more suitable for processing images from irregular bodies.

We use *ISIS3* to construct the required data products (Section 3). Camera models for MapCam, PolyCam, and SamCam are presently included in *ISIS3* (Golish et al., 2018). To advance small-body image processing, specific *ISIS3* applications were also augmented. In particular, we incorporated new ray-tracing libraries*,* which perform more accurate and efficient image registration of data from irregular planetary shapes. In turn, the increased efficiency introduced by these libraries allowed us to improve image orthorectification techniques.

Additionally, we modified the *ISIS3* bundle adjustment application (*jigsaw*, Edmundson et al., 2012, Section 5.3) to solve for surface points in either *x, y, z* (Cartesian) or longitude, latitude, radius (planetocentric) coordinates. This is an option presented to the user in the *jigsaw* interface. Previously, *jigsaw* was limited to solving for surface points in latitude, longitude, radius coordinates. These tools, and in particular the use of Cartesian coordinates, facilitate the characterization of irregular bodies with a Digital Shape Kernel (DSK), a 3D tessellated shape model format recently introduced by NASA's Navigation and Ancillary Information Facility (NAIF; Acton et al., 2017). These additions enhance our ability to generate accurate raster basemaps of Bennu. Nonetheless, traditional 2D mosaics often fall short when representing the surface of irregular small bodies. Consequently, we have developed tools that are able to export the results from *ISIS3* so that they can be viewed in 3D.

5.2 Tessellated Shape Model and Ray Tracing Support

Reference ellipsoids, often defined in a SPICE planetary constants kernel (PCK), are a simple way to describe the shape of a planetary body (Acton et al., 2017). If the reference ellipsoid of an irregular body deviates too significantly from the true shape it will not be useful for image registration and mapping. Similarly, while an *ISIS3* Digital Elevation Model (DEM) provides higher resolution information that may account for local terrain, these elevation values are mapped in a 2D digital representation in terms of latitude and longitude relative to an ellipsoid. Consequently, overhanging terrain is not fully represented. These shortcomings are addressed when the shape of a planetary body is described by a 3D tessellated plate model, such as a SPICE DSK, which can represent a shape of arbitrary complexity. In tandem with a realistic camera model, a DSK can facilitate accurate projection of data onto a planetary body by more precisely describing the true shape of that body. This capability improves the estimates of the pixel ground sample distance for images, which are calculated through ray tracing. For OCAMS data from asteroid Bennu, we exclusively use a tessellated shape model to perform digital mapping.

In *ISIS3*, images can be projected to a DSK using two different ray-tracing techniques. Forward-driven ray tracing is used for determining image backplanes that provide the photometric angles, body-fixed rectangular, and body-fixed geographic coordinates on a per pixel basis. Backplanes are a key input to the 3D point-cloud mosaicking process (Section 5.3) and photometric modeling as described in Golish et al., (2018). Backward-driven ray tracing is preferred for conventional 2D mosaicking, which depends on the orthorectification and cartographic projection of individual image frames. The backward-driven approach is favored for its efficiency, as described later in this section.

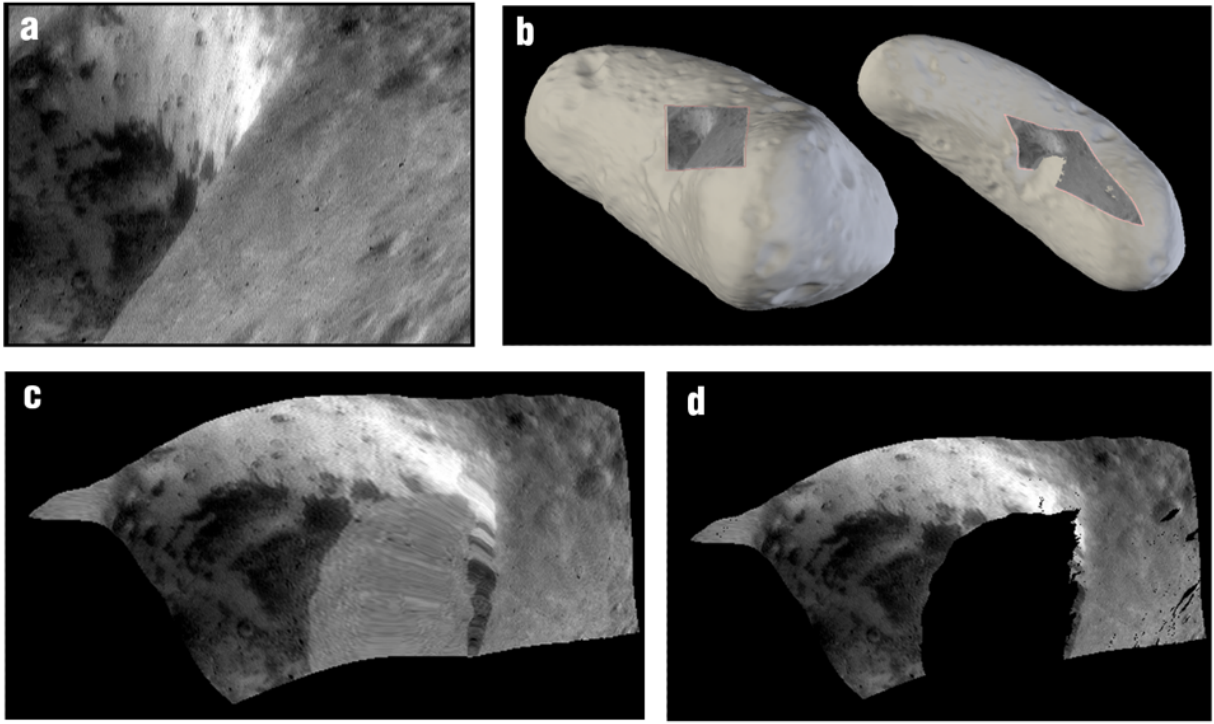

**Figure 6.** (a) A raw, unprojected NEAR Multi-Spectral Imager frame of Psyche Crater on asteroid (433) Eros (M0132393443F4_2P_IOF_DBL.FIT) which is available through the Planetary Data System (PDS). (b) The same image displayed on the Eros shape model in the Small Body Mapping Tool (SBMT); viewing the image at different perspectives reveals a significant amount of occluded terrain between the fore- and background. (c) The same image, map-projected in simple cylindrical using the original 3D shape model projection technique in *ISIS3*; this technique is not able to handle terrain occlusion and erroneously interpolates data into the occluded area using pixels from surrounding terrains. (d) The same image map-projected using the more robust Bullet ray tracing engine; this technique more accurately distinguishes between visible and occluded terrain and culls the occluded region in the resulting map-projected image.

To improve the performance and accuracy of cartographic image processing from irregular bodies, two new ray-tracing software libraries have been incorporated into *ISIS3*. Prior to this, *ISIS3* used the NAIF DSK toolkit to provide tessellated 3D shape model support. The new libraries are Embree, a ray tracing engine developed at Intel (Wald et al., 2014), and Bullet, a collision detection and physics engine created by Coumans (2010). Both libraries have robust ray-tracing support and demonstrate success in environments that require both accuracy and efficiency. In *ISIS3* the Embree and Bullet implementations can read the contents of a DSK file (vectors and indexed triangular plate descriptors) and render them into internal representations that are optimized for each library. The Embree system offers high-performance ray-tracing kernels that are optimized for the latest Intel processors. Despite its computational efficiency, the Embree engine uses single-precision floating-point geometry representations, which can lead to precision issues in the *ISIS3* environment. This issue is especially apparent when the resolution of the image is significantly different than that of the shape model and that difference approaches the 32-bit single-precision floating-point limit. The Bullet Continuous Collision Detection and Physics Library ("Bullet") was identified as a replacement for the NAIF DSK toolkit because it enables double-precision geometry and ray-tracing calculations. It also provides GPU support, robust real-time collision detection and multi-physics simulations. This extensibility made Bullet a particularly appealing choice and will likely facilitate future innovations. The improved efficiency of the new libraries allows us to advance the backward-driven ray tracing techniques that support image orthorectification.

Traditionally, the creation of 2D cartographic products in *ISIS3* is a backward-driven process (Anderson, 2013). Starting with a 2D map projection (e.g., simple cylindrical, sinusoidal), backward-driven orthorectification computes image coordinates at an observation from a latitude/longitude (map) coordinate. Working backwards from this point, each unique projection equation is solved through a camera model and a line/sample (image) coordinate is returned. A single orthopixel at the coordinate is subsequently computed using an interpolation technique (e.g., bilinear, cubic convolution).

In contrast, forward-driven orthorectification starts from a line/sample coordinate in an image and works through a camera model to compute the ground coordinate on a shape model. A 2D map projection equation is separately applied to that ground point to compute the corresponding latitude/longitude (map) coordinate. The computational complexity of forward-driven orthorectification makes this approach less efficient than the backward-driven method. Moreover, the backward model typically selects a subset of pixels from an image to produce each map pixel, whereas the forward model will project all pixels in an image. Hence, in *ISIS3* the backward projection is favored for orthorectifying images.

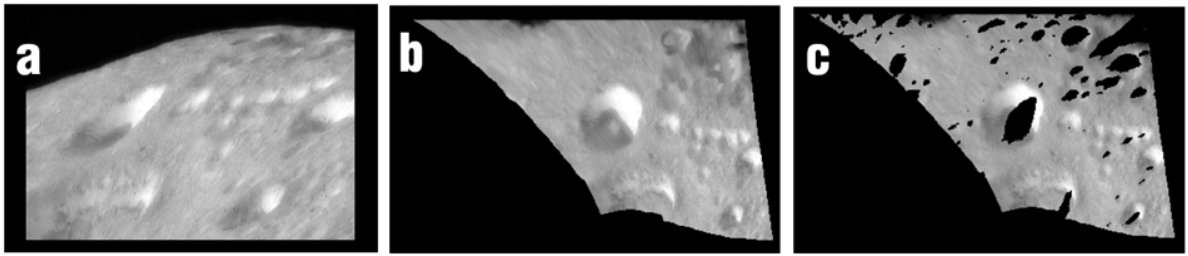

**Figure 7.** (a) A raw, unprojected NEAR Multi-Spectral Imager frame capturing terrain from an oblique view that includes the limb of the asteroid (433) Eros (M0132382244F4_2P_IOF_DBL.FIT). (b) The same image, map-projected in simple cylindrical using the original 3D shape model projection technique in *ISIS3*. (c) The same image map-projected using the more robust Bullet ray tracing engine. Note the effect of interpolation when occluded data are not properly defined and culled. Unlike the obvious interpolation effects seen in Fig. 5c, these errors are significant but not obvious. Misleading and invalid data are introduced at the bottom of crater floors in (b), which are clearly not observed in the raw frame (a). The impact of this interpolation is difficult to assess until the occluded pixels are removed (c).

One complication of backward projection is identifying areas of terrain occlusion. Occlusion events occur when terrain in the foreground of an image partially blocks scenery that is further in the depth of field (Figure 6a). In planetary mission data, this is commonly seen at the limb and in oblique imagery due to surface features such as boulders or crater edges. Even when observations are well planned, irregular planetary bodies that have significant concavities (e.g., Psyche Crater on Eros), elongated axes (e.g., Eros, Itokawa), or interconnected lobes (e.g., 67P/Churyumov–Gerasimenko) are especially prone to terrain occlusion. Thus, any software responsible for projecting images onto a shape model must determine where occluded terrain occurs and cull it. Prior to the work we present here, *ISIS3* interpolated across missing occluded terrain when orthorectifying images. In large areas of occlusion, this interpolation introduced significant regions of obvious, invalid data into map projected products (Figure 6c). In smaller areas of occlusion, we observe that these effects can result in misleading data that appear realistic but are invalid (Figure 7).

In *ISIS3*, recognizing and culling occlusion has been improved by incorporating additional ray-tracing tests that upgrade the backward-driven orthorectification algorithm. The geometries that describe these ray-tracing tests are depicted in Figure 8. Methods for dealing with occluded terrain differ for the backward- and forward-driven ray-tracing techniques. In the forward approach, where the ray computation begins at the camera (Figure 8, Point A), the camera views the first piece of terrain along a ray and stops. However, in backward projection, where a ray begins on the surface, ray tracing must continue to determine where other parts of the surface obstruct the camera view (Figure 8, Point B).

Using a backward-driven algorithm, the pixel value of a latitude/longitude (map) coordinate that is calculated from a particular line/sample (image) coordinate is determined through a 2D map projection. Once the map pixel is defined (Figure 8, Point B), the vector between this latitude/longitude (map) coordinate and the center of the body can be computed in rectangular coordinates $(x, y, z)$. The ray describing this vector is extended above the surface to a terminal point that is beyond the maximum radius of the body. This serves as a virtual observer point; the ray direction can be reversed from this point to provide the desired nadir view of the surface used in orthorectification. From the observer point, a ray intersection with the shape model (DSK) is calculated to identify the ground point (triangular plate) at the surface. The triangle of intersection is always the one closest to the observer for ray tracing operations. It is provided to the camera model to compute the image coordinates and solve for the backward projection. The validity of the computed ground point is verified by computing the emission angle, i.e., the angle between the ray from the surface intersect point to the instrument coordinate and the normal vector of the intersected triangle. If this angle is greater than 90°, the ground point is not visible from the instrument and the pixel is set to null. If the emission angle is less than 90°, the ground point is valid and a corresponding orthomap pixel is computed using interpolation.

To ensure that the view of the ground point is not occluded by other terrain, a second ray is also traced. Forward projection is used to trace along the ray that is calculated between the ground point and image focal plane by the backward projection. The forward projection then calculates the triangle of intersection on the shape model (DSK). If the same triangle is not intersected by both the forward and backward traced rays, then this test fails and the ground point is deemed occluded. To further control the accuracy of the test, a tolerance on the Euclidean distance between backward and forward intersection points can be specified on the triangle. Because the computational overhead of the forward projection process nearly doubles the already expensive process of the backward projection, this test is only invoked when using either the Embree or Bullet ray tracing engines in *ISIS3*.

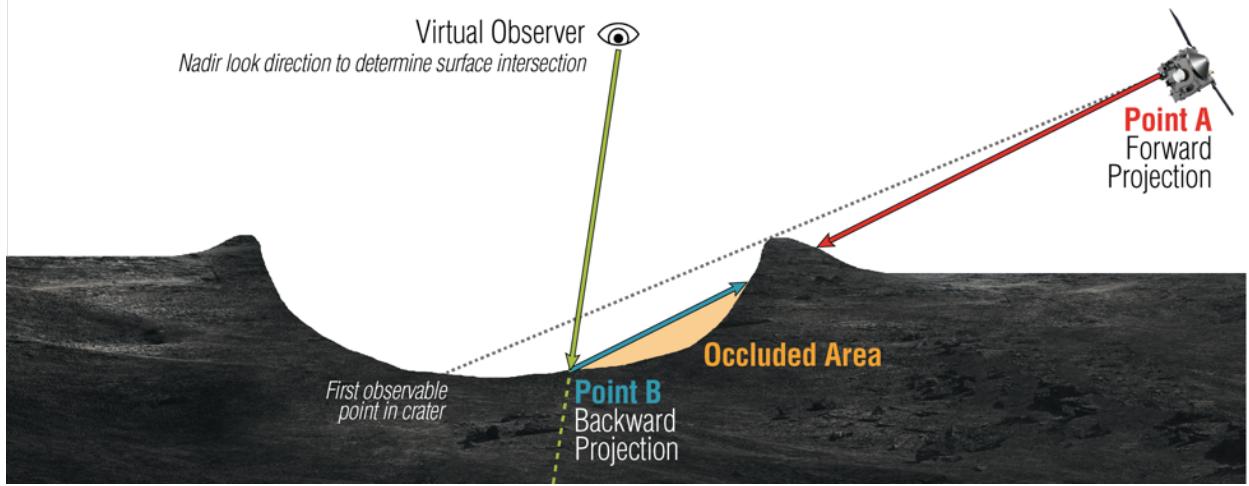

**Figure 8.** A cartoon illustrating the rays traced from a desired location on the surface of Bennu using forward projection (Point A, red) and backward projection (Point B, blue). The position of the virtual observer that is used during the orthorectification of the same point on the surface is also depicted. Note that in this diagram, the ray from the forward and backward projections do not intersect at the same place on the surface, indicating that occluded terrain is present.

5.3 Mosaicking Images from Small Irregular Bodies

To provide an initial "quick look" of Bennu shortly after data downlink, image mosaics are developed by projecting images to a DSK using the OCAMS camera model and early estimates of observation geometry that are provided by predicted SPICE kernels. These mosaics are considered uncontrolled because the images are reprojected using preliminary instrument pointing and spacecraft trajectory information. Despite the utility of their rapid production, these products are not suitable for mission critical planning due to errors in image geometry.

Accurately mapping the position and dimensions of surface hazards on Bennu is a vital step towards identifying candidate sample-sites. Hence, the basemaps used by the OSIRIS-REx team to derive higher-level data products are controlled. In the control process, images are registered to each other through the measurement of common features. The error budget associated with image-based photogrammetric control is well-established from its decades of use in engineering and remote-sensing (e.g., Granshaw, 1980; Toutin, 2003; Edmundson et al., 2012; Becker et al., 2015; Henriksen et al., 2017). Following the established *ISIS3* photogrammetric control procedure, errors most commonly originate from the following sources: 1) sensor calibration and digital camera model errors, such as uncharacterized geometric distortion or thermal effects; 2) Errors in SPICE data for spacecraft pointing and position at the time of exposure; 3) Image matching errors and measurements of tie points; 4) Misregistration between images and terrain; and 5) Errors in the triangulated surface points that result from the accumulation of errors propagated into the bundle adjustment process (*jigsaw* in *ISIS3*).

To mitigate these errors, we follow several recommended approaches. For every filter and focal length of the three OCAMS imagers, we account for geometric distortion (Pelgrift et al., 2018) within their respective *ISIS3* camera model. We also use reconstructed SPK and CK SPICE kernels to provide the best available estimates of a priori spacecraft pointing and position during image acquisition.

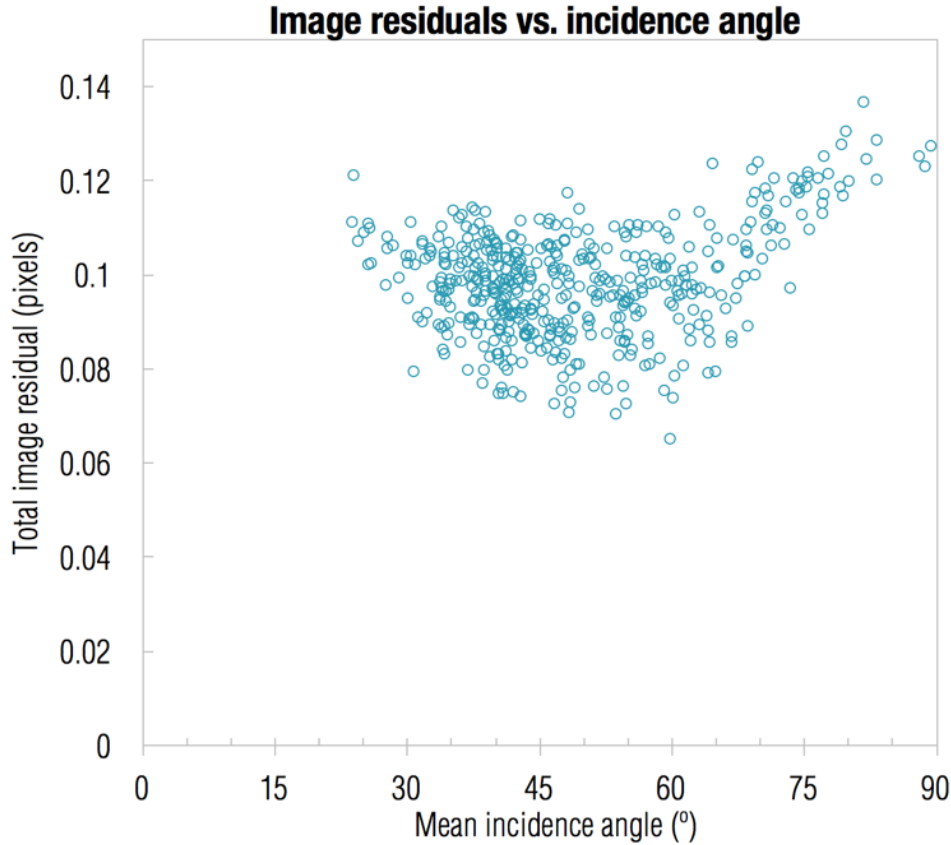

**Figure 9.** A plot of the residuals from a *jigsaw* bundle adjustment versus mean incidence angle for individual simulated OCAMS images from the Baseball Diamond Panchromatic Strategy. Simulated data were created using techniques similar to those described by Golish et al., (2018) and mosaicked according to the procedure described by Becker et al., (2017). The plot illustrates a trend commonly seen in these simulations, where residuals <0.1 pixel correspond to images acquired at incidence angles between approximately 25º-75º.

However, image-to-image misregistration typically persists even with use of reconstructed kernels. Thus, the creation of either manual or automated image registration is required to further minimize reprojection errors (< 0.6 average pixel residual error). In the control process, images are registered to each other through the measurement of features, known as tie points, common to overlapping images. Images may also be registered to the ground by identifying corresponding features between them and existing base maps or shape models. These features are control points. A control network is comprised of the 3D coordinates and image measurements of tie and control points.

We generate tie-points through automated image-matching and outlier rejection schemes that were initially developed for the MESSENGER mission (Becker et al., 2017). These automated techniques perform best when images are acquired under uniform and non-extreme photometric conditions. As with SPG, incidence angle plays a particularly important role in image-matching, with mid-range values yielding the lowest residual error ("residuals") in triangulated surface points (Figure 9). Hence, data products that require high geometric (rather than spectrophotometric) fidelity may benefit from images acquired under those illumination conditions.

One notable difference between the standard processing techniques used for controlling images from large planetary targets vs. small bodies is the increased importance of accurately tying images to the ground. Irregular small bodies often have more dramatic global terrain variations than their larger, more spheroidal counterparts. Thus, it is critical that images are accurately tied to the ground so that rays can be correctly traced between a shape/terrain model and spacecraft instrument (Section 5.2). Inaccurate ray tracing will lead to false detections of occluded terrain, incorrect pixel scale calculations, and result in persistent image-to-image misregistration. To avoid these issues, we have developed two additional approaches for tying images to the ground. In the first, we create simulated images of the surface using a DSK and a photometric model (Golish et al., 2018). Control point measurements are then obtained between real and simulated images. This is analogous to controlling to a terrain shaded relief map. However, because illumination conditions can be more precisely controlled when creating these simulated images (particularly for irregular bodies) they may offer better results. In the second approach, we obtain ground control measurements by using the control network and image geometry provided by the SPC process (Le Corre et al., 2017).

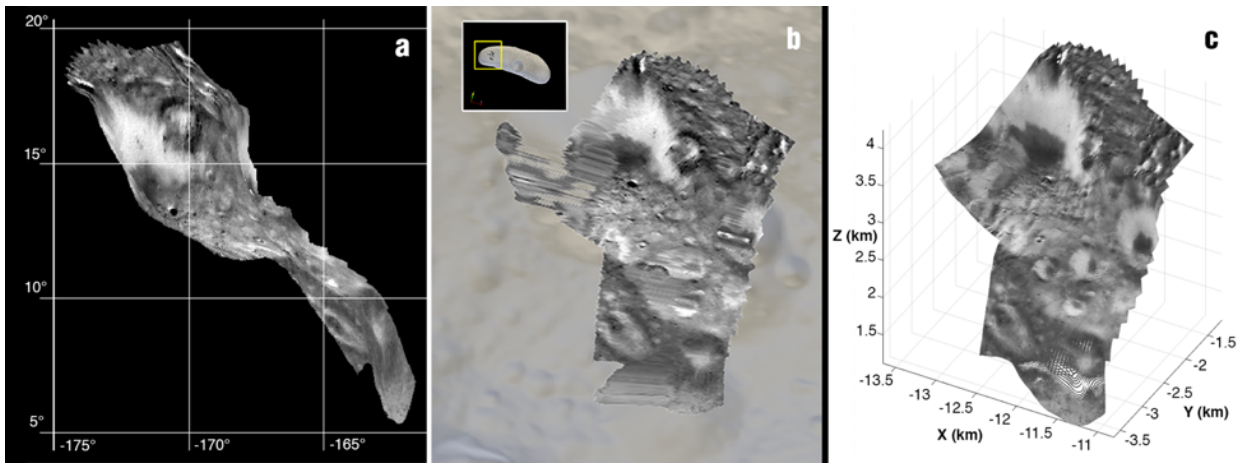

**Figure 10.** (a-c) A set of 14 NEAR-Shoemaker Multi-Spectral images (~4.5 m pixel scale) taken near the nose of asteroid (433) Eros. (a) Here the data are mapped in a simple cylindrical projection centered at 0,0 (at present time simple cylindrical is the only map-projection supported by SBMT). Before being mosaicked, these data were projected to a 3D tessellated shape model of (433) Eros using the Bullet ray tracing engine. (b) The mosaic from the panel (a) texture mapped back on to the same shape model of (433) Eros using SBMT. Note the loss of 3D information that occurs when the image mosaic is transformed into a 2D map projection, which results in a distortion when the data is re-mapped back to 3D. We note that other 2D map projections (e.g., sinusoidal, azimuthal, etc.) illustrate similar data loss, which is likely due to the elongated shape of Eros and the concave nature of the terrain depicted. (c) The same images mosaicked as a Cartesian point cloud. This technique combines the results of the *ISIS3* bundle adjustment without a 2D map projection, and allows accurate mapping of pixels without any data loss, or subsequent distortion. Note that the point cloud appears sparser where less or lower resolution data are present. Future work will involve implementing schemes that interpolate these point-clouds into a triangulated irregular network (TIN).

Tie and control point coordinates and image measurements serve as input to the next step in the control process, the least-squares bundle adjustment (Brown, 1976). The bundle adjustment (*jigsaw*) improves image position and pointing parameters and generates the triangulated ground coordinates of tie and control points. As mentioned earlier (Section 5.1), we have modified *jigsaw* to generate surface points in either *x, y, z* (Cartesian) or longitude, latitude, radius (planetocentric) coordinates. Again, Cartesian coordinates are better suited to uniquely define feature locations on irregularly shaped bodies. The error propagation option in *jigsaw* provides uncertainty values for each of the parameters in the bundle adjustment, as calculated by the covariance matrix.

With updated image geometry from the bundle adjustment, images are photometrically corrected to predefined standard illumination conditions to remove photometric effects and minimize the presence of seams. For 2D mosaics, individual images are translated into a common cartographic projection before they are mosaicked. During map projection, the frames are corrected for lens distortion while being rescaled to a uniform resolution. This minimizes the number of times that the image is resampled and ensures that the control network is created with unaltered pixel data. Accordingly, we do not correct for geometric distortion in previous processing steps. The images are map projected using their updated geometry and a 3D shape model to provide accurate registration to the body and remove topographic effects. During the creation of the mosaic, the determination of which images are "on top" (i.e. visible) can be based on a number of parameters, such as resolution, photometric angles or average values of reflectance, or simply acquisition time (e.g. newer images are added on top of the mosaic).

For 3D point cloud mosaics, the improved geometry from the bundle adjustment is used to recompute image backplanes in *ISIS3*, including the body-fixed rectangular values (*x, y, z*) per pixel. Photometrically corrected images and their updated backplanes are then exported from *ISIS3* as a dense point cloud (Figure 10c) with the appropriate reflectance values assigned to those points. These point clouds can be visualized using a number of different tools, such as the Point Cloud Library (PCL; Aldoma et al., 2012) or MATLAB's Computer Vision System Toolbox (MathWorks, 2017). Future work will involve implementing schemes that interpolate the point-cloud into a TIN.

5.4 Displaying Mosaics and Orthoimages from Irregular Bodies

Conventional 2D Geographic Information System (GIS) software, especially packages that assume the uniqueness of latitude and longitude, are insufficient for visualizing data from irregular small bodies. Even when a 3D shape model is used to accurately project images, most standard cartographic projections are based on spherical objects and fail to accurately portray data from these bodies. Nevertheless, 2D basemaps of Bennu are still generated using standard map projections because global visualization in 3D often requires manual manipulation through a user-interface to rotate a data set, resulting in examination in a piecemeal fashion. This can lead to a more gradual discovery of geographic trends. In contrast, a 2D map will convey a global data set in a single visual representation, aiding in the ability to readily interpret geospatial relationships. Hence, both 2D and 3D map products of Bennu are capable of benefiting the OSIRIS-REx mission site-selection process.

Map projections that have been optimized for small bodies with elongated axes (e.g., upturned ellipsoid transverse azimuthal; Fleis et al., 2007) were recently implemented in *ISIS3* and can be used when appropriate. Otherwise, 2D basemaps are primarily displayed using standard map projections in the JAsteroid tool customized for the OSIRIS-REx mission (Smith et al., 2015). When images are registered to a 3D shape model in *ISIS3*, the resulting 2D mosaics can be accurately texture-mapped back on a 3D shape model. This is straightforward to accomplish in a tool such as the Applied Physics Laboratory's Small Body Mapping Tool (SBMT; Barnouin et al., 2018). SBMT is able to display basemaps draped upon a 3D shape model provided they are in simple cylindrical projection and the appropriate metadata are supplied. For areas without overhanging terrain, this scheme works well; however, limitations exist when working with images of a surface that is not uniquely described by latitude and longitude or across elongated axes. If data are lost when an image is translated from camera space into a 2D map, then these data cannot be recovered by simply re-applying them to a 3D model. In this case, we recommend exporting the results of *ISIS3* as a Cartesian point cloud (Section 5.3) that can be visualized as-is, or interpolated into a TIN of the surface. This provides visualization of image mosaics free from the distortions that are introduced during the 2D map projection (Figure 10). Point clouds and TINs may not supplant the utility of a 2D basemap for every small body, but they do offer additional context and complement traditional raster products.

## 6 Conclusions

Working backward from the foundational products critical to the success of the OSIRIS-REx mission, we used a systems engineering approach to identify suitable imaging strategies, plan observational campaigns, and assess the necessary techniques to perform image-based mapping of a small body. We present the resulting imaging strategies and observational plans, which are optimized for specific data products that fulfill mission science requirements.

Additionally, we describe the current best practices to address the unique challenges associated with processing image data of irregularly shaped small bodies. Emphasis is placed on the use of a camera model in tandem with an accurate 3D tessellated shape model for describing the shape of small planetary bodies. Likewise, the ability to solve for camera geometries and display mosaics in body-fixed Cartesian coordinates (*x, y, z*) is recommended for generating basemaps that correctly illustrate the spatial relationships of an irregular surface. This is especially important for overhanging terrains that cannot be uniquely described by latitude and longitude. The approach, methodologies, and best practices presented here aid the OSIRIS-REx mission in its primary objective of acquiring a sample of Bennu and are likely to benefit future missions to small body targets.

## Acknowledgments and Data

This material is based upon work supported by NASA under Contract NNM10AA11C issued through the New Frontiers Program. Dante Lauretta of the University of Arizona, Tucson, is the principal investigator, and the University of Arizona also leads the science team and the mission's science observation planning and data processing. NASA's Goddard Space Flight Center in Greenbelt, MD, provides overall mission management, systems engineering and the safety and mission assurance for OSIRIS-REx. Lockheed Martin Space in Littleton, CO built the spacecraft and is providing flight operations. Goddard and KinetX Aerospace are responsible for spacecraft navigation. OSIRIS-REx is the third mission in NASA's New Frontiers Program. NASA's Marshall Space Flight Center in Huntsville, Alabama, manages the agency's New Frontiers Program for the Science Mission Directorate in Washington D.C. The Earth Gravity Assist OCAMS data used in this work are available through the Planetary Data System (http://pds.nasa.gov/). The Bullet and Embree libraries are included in publically available releases of ISIS3 as of version 3.5.2 and later (https://github.com/USGS-Astrogeology/ISIS3). The body-fixed rectangular features included in the *jigsaw* application will be made available in version 3.6 of *ISIS3*. NEAR Multi-Spectral Imager data of asteroid (433) Eros used to demonstrate of software capabilities are available through the Planetary Data System Small Bodies Node (https://pds-smallbodies.astro.umd.edu/).

**Abbreviations Table**

| | |
|---|---|
| BRDF | Bidirectional Reflectance Distribution Function |
| DEM | Digital Elevation Model |
| DSK | Digital Shape Kernel |
| DTM | Digital Terrain Model |
| ECAS | Eight Color Asteroid Survey |
| FB | Flyby |
| FOV | Field of view |
| GIS | Geographic Information System |
| GM | Product of the gravitational constant (G) and mass of a planetary body (M) |
| GNC | Guidance Navigation and Control |
| GPU | Graphics Processing Unit |
| ISIS3 | Integrated Software for Imagers and Spectrometers version 3 |
| JMARS | Java Mission-planning and Analysis for Remote Sensing |
| LPI | Lunar Planetary Institute |
| MapCam | Medium-resolution imager in OCAMS |
| NAIF | NASA's Navigation and Ancillary Information Facility |
| NEAR | Near Earth Asteroid Rendezvous |
| OCAMS | OSIRIS-REx Camera Suite |
| OLA | OSIRIS-REx Laser Altimeter |
| OSIRIS-REx | NASA's Origins, Spectral Interpretation, Resource Identification, Security-Regolith Explorer mission |
| OTES | OSIRIS-REx Thermal Emission Spectrometer |
| OVIRS | OSIRIS-REx Visible and InfraRed Spectrometer |
| PCA | Principal Component Analysis |
| PCK | Planetary constants kernel |
| PDS | Planetary Data System |
| PolyCam | High-resolution imager in OCAMS |
| PSFD | Particle Size Frequency Distribution |
| RBD | Relative Band Depth |

| REXIS | REgolith X-ray Imaging Spectrometer |
|-------|--------|
| RGB | Red, green, blue color model |
| SamCam | Asteroid sampling context imager in OCAMS |
| SBMT | Small Body Mapping Tool |
| SNR | Signal-to-noise ratio |
| SOCET SET | SOftCopy Exploitation Toolkit, a stereophotogrammetry software platform from BAE Systems, Inc. |
| SPC | Stereophotoclinometry |
| SPG | Stereophotogrammetry |
| SPICE | Spacecraft, Planet, Instrument, C-matriX, Events Toolkit developed by NASA's Navigation and Ancillary Information Facility |
| SRC | Sample Return Capsule |
| TAG | Touch-and-Go |
| TAGSAM | Touch-and-Go Sample Acquisition Mechanism |